\documentclass[prb,twocolumn,amsmath,amssymb,citeautoscript]{revtex4}

\usepackage{graphicx}
\usepackage{dcolumn}
\usepackage{bm}
\usepackage{bbm}
\usepackage{dsfont}
\usepackage[bbgreekl]{mathbbol}

\newcommand{\bq}{\begin{equation}}
\newcommand{\eq}{\end{equation}}

\newcommand{\bqa}{\begin{eqnarray}}
\newcommand{\eqa}{\end{eqnarray}}

\DeclareSymbolFont{mybbsl}{U}{bbm}{m}{sl}
\DeclareMathSymbol{\complext}{\mathord}{mybbsl}{"74}
\DeclareMathSymbol{\TimeOrderT}{\mathord}{mybbsl}{"54}
\DeclareMathSymbol{\complexG}{\mathord}{mybbsl}{"47}

\begin{document}

\title{Long tunneling contact as a probe of fractional quantum Hall
neutral edge modes}

\author{B. J. Overbosch$^{1}$}
\author{Claudio Chamon$^{2}$}

\affiliation{$^{1}$\!\!\!
Physics Department, Massachusetts Institute of Technology, Cambridge, MA 02139, USA
\\
$^{2}$\!\!\!
Physics Department, Boston University, Boston, MA 02215, USA
}

\date{October 7, 2008; v2 July 9, 2009}

\begin{abstract}
We study the tunneling current between edge states of quantum Hall
liquids across a single long contact region, and predict a resonance
at a bias voltage set by the scale of the edge velocity. For typical
devices and edge velocities associated with charged modes, this
resonance occurs outside the physically accessible bias
domain. However, for edge states that are expected to support neutral
modes, such as the $\nu=\frac{2}{3}$, and $\nu=\frac{5}{2}$ Pfaffian
and anti-Pfaffian states, the neutral velocity can be orders of
magnitude smaller than the charged mode and if so the resonance would
be accessible. Therefore, such long tunneling contacts can resolve the
presence of neutral edge modes in certain quantum Hall liquids.
\end{abstract}

\maketitle

\section{Introduction}

Quantum Hall (QH) states are incompressible quantum fluids where all
bulk excitations are gapped, but gapless modes exist at the
boundaries. In the integer effect, edge states can be understood in a
simple way for non-interacting electrons~\cite{Halperin}, with an edge
channel matching each filled Landau level in the bulk, as the Landau
bands bend at the edges of the system due to the confining potential
and cross the Fermi level. In the fractional effect the situation is
richer, and there is a one-to-one relation due to gauge invariance
that ties the bulk states, classified by 2+1D Chern-Simons theories,
and the gapless edge modes~\cite{Wen}. Depending on the bulk filling
fraction or the details of edge confinement, the edge theory may
contain, in addition to a charge mode that carries the quantized Hall
currents, neutral modes. For example, even for a $\nu=1$ QH state,
neutral modes are present if the edge is smooth or
reconstructed~\cite{Chamon+Wen1994}. For fractional QH states, even
for sharply defined edges, neutral modes may be present. Such is the
case for $\nu=\frac{2}{3}$ states~\cite{MacDonald,Kane-etal}, as well as for the
$\nu=\frac{5}{2}$ Pfaffian and anti-Pfaffian non-Abelian
states~\cite{Pfaffian,anti-Pfaffian}, {and~the~situation~becomes~even~%
richer} if the edges of such states undergo
reconstructions~\cite{Overbosch+Wen2008}.

Chiral charge modes, which cannot be localized by disorder, are
closely tied to the quantization of the Hall conductance; hence the
existence of these modes is unavoidable. Experiments have been
designed to probe the propagation of these charge modes, in particular
to measure their wave
velocity~\cite{Zhitenev-etal,Ashoori-etal}. On the other hand, to
the best of our knowledge, there has not been any experimental result
that confirms the existence of the neutral modes.

There are a number of reasons as to why one should seriously look into
ways of detecting neutral edge modes. For example, there are
theoretically unresolved experimental findings on tunneling on the
edges of QH liquids in cleaved-edge overgrown
samples~\cite{Chang-etal,Grayson-etal} which could be better
understood if information on the neutral modes were available. More
specifically, in these experiments one measures a non-linear $I-V$
characteristic of Luttinger liquid behavior at the edges; however, the
power-law exponent is not in agreement with theoretical
predictions~\cite{Wen,Shytov-etal}. Instead, these exponents match
those obtained if one had only the charge mode and no neutral
ones. However, one cannot construct an operator for creating an
excitation with the charge of an electron and fermionic statistics
using the charge mode alone. Hence, the neutral modes are both the
champions and the villains lingering over the resolution of this
puzzle, and this has led to proposals that the neutral modes may be
either extremely slow~\cite{Lee+Wen} or topological and
non-propagating at all~\cite{Fradkin+Lopez}. Another reason to probe
neutral modes is that, in the case of the interesting non-Abelian
states, these are the modes that carry the non-local information of
the order and twinning of edge quasiparticles.

\begin{figure}
\includegraphics[width=3.375in]{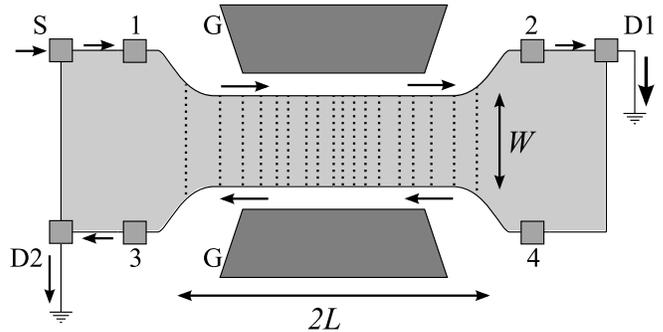}
\caption{Tunneling between two edges in a quantum long contact (QLC)
does not occur at a single site, but rather over a range of positions
along the edge. 
Arrows indicate propagation direction of current.
\label{fig:longQPC}}
\end{figure}

The objective of this paper is to propose a way to probe neutral 
edge
modes. The proposed set-up consists of a long contact region,
or quantum long contact (QLC),
 in which
there are several interfering paths for tunneling charge from two
opposite edges of a Hall bar, as depicted in Fig.~\ref{fig:longQPC},
resembling an AC Josephson junction. The idea of exploring
interference between tunneling paths is reminiscent of a two-point
contact interferometer~\cite{CFKSW9731} (2PC) for probing quasiparticle
statistics.
Both methods are sensitive to neutral edge modes,  the main difference being the observation window: the long contact setup probes slower edge velocities than the two-point contact setup.

We find that coherent tunneling inside a QLC gives rise to a resonance in the tunneling current at zero
temperature for a bias voltage $V_\text{res}$ given by
\begin{align}
\frac{eV_\text{res}}{\hbar}
=\frac{vW}{\ell_B^2},
\label{eq:res}
\end{align}
where $W$ is the width of the QLC, $\ell_B$ is the magnetic length, and
$v$ is the slowest edge velocity associated with the tunneling
quasiparticle. The origin for the resonance has a simple
explanation. 
The interference of a tunneling quasiparticle between two paths, separated by a distance $x$ along the edge, is guided by two phases:
 on the
one hand there is the Aharonov-Bohm phase $(e^*/e)\, x W/\ell_B^2$ that
basically multiplies the quasiparticle charge $e^*$ with the flux
enclosed in the area $W x$; on the other hand there is the phase
$\omega_J t$ that is introduced by an applied bias voltage $V$ between the two edges, with
Josephson frequency $\omega_J=e^* V/\hbar$ and $t=x/v$.
 The resonance occurs at the stated voltage $V_\text{res}$ when
the two phases become equal and give rise to constructive
interference. The resonance condition follows from the interference
among multiple tunneling paths along the length $L$ of the QLC;
however, notice that the length of the channel drops out of the
resonance condition Eq.~(\ref{eq:res}). The resonance becomes sharper
for longer lengths $L$ of the QLC. At finite temperature $T$ the
resonance will be reduced and for temperature $2\pi \,T>e^*
V_\text{res}$ it will be washed out.
A sharp resonance in the tunneling current will lead to a
strong peak followed by a strong dip in the tunneling conductance at
non-zero bias.

Now, if there are multiple edge velocities associated with propagation
of the quasiparticle along the edge, there are in principle multiple
phases $\omega_J x/v_i$, one for each velocity. We are especially
interested in a situation where there are two velocities: one fast
velocity associated with the charged mode, and one slow velocity
associated with the neutral mode(s). For the charged mode the edge
velocity is expected on general grounds to be determined by the scale
set by electron-electron interactions, $v_c\sim
(e^2/\epsilon)/\hbar=\alpha \, c/\epsilon$, where $\epsilon$ is the
dielectric constant of the medium ($\epsilon_{\rm GaAs}\approx
12.9$). Therefore, the charge mode velocity is of order $\sim 10^5
\text{m/s}$. With a width $W\sim 10\ell_B$
and $\ell_B\sim 10 \text{nm}$ we would find $V_\text{res}\sim 0.1
\text{V}\simeq 10^3 \text{K}$; the current that would have to be driven through the sample
at such a voltage would surely destroy the quantum Hall state. A resonance due to
such a fast velocity is thus not likely experimentally accessible at a QLC. A
neutral mode velocity is not bound to the scale set by Coulomb interactions 
though, and can in principle be orders of magnitude smaller.

We proceed in Sec.~\ref{sec2} with a detailed calculation of the tunneling current to determine the precise lineshape of the resonance; Fig.~\ref{fig:Gtunplots} illustrates the main result of this paper. In Sec.~\ref{sec3} we focus on the range of accessible slow edge velocities and compare the observation ranges of the QLC and the 2PC. We conclude in Sec.~\ref{sec4}.

\section{Tunneling current through a quantum long contact\label{sec2}}

In this section we calculate the tunneling current through a QLC  to determine the lineshape of the resonance as a function of bias, temperature, tunneling exponent and edge velocity.
 The tunneling current due to $N$ (discrete) tunneling sites was
calculated in linear response in Ref.{~}{\onlinecite{CFKSW9731}},
\begin{multline}
I_\text{tun}(\omega_J)=
e^*\sum_{i,j=1}^N\frac{\Gamma_i\Gamma^*_j+\Gamma_i^*\Gamma_j}{2}
\int_{-\infty}^\infty dt \,e^{i\omega_J t}\times\\
P_{g\over 2}(t+x_{ij}/v)
\;
P_{g\over 2}(t-x_{ij}/v)-(\omega_J\leftrightarrow -\omega_J).
\label{eq:ItunCFKSW}
\end{multline}
Here $x_{ij}=x_i-x_j$ and edge quasiparticle propagator $P_{g/2}(t)$ is given
by
\begin{gather}
P_{g\over 2}(t)=\begin{cases}
\displaystyle\frac{1}{(\delta+it)^g}& \text{for $T=0$, $\delta=0^+$},\\
\displaystyle\frac{(\pi T)^g}{(\delta+i\sinh\pi T t)^g}& \text{for $T\ne0$}.
\end{cases}
\end{gather}
In this paper we will generalize Eq.~(\ref{eq:ItunCFKSW}) by
making the discrete number of tunneling sites into a continuous
distribution, $\Gamma_i\to\gamma(x)$, and to separate contributions
from charged and neutral modes, which come with distinct edge
velocities $v_{c/n}$ and tunneling exponents $g_{c/n}$,
\begin{widetext}
\begin{multline}
I_\text{tun}(\omega_J)=
e^*\int\!\! dx dy \: \gamma(x)\,\gamma^*(y)\int_{-\infty}^\infty
\!\!\!\!dt\,e^{i\omega_J t}
\;P_{g_c\over 2}\left(t+\frac{x-y}{v_c}\right)
P_{g_c\over 2}\left(t-\frac{x-y}{v_c}\right)
P_{g_n\over 2}\left(t+\frac{x-y}{v_n}\right)
P_{g_n\over 2}\left(t-\frac{x-y}{v_n}\right)
\\-(\omega_J\leftrightarrow -\omega_J).
\label{eq:Itun-conntinuum}
\end{multline}
\end{widetext}
See Fig.~\ref{fig:longQPC} for a sketch of the setup. 
We assume that the entire bulk has the
same filling fraction, and the edges are the modes associated with
that bulk state. In the narrow region under the QLC we do not allow bulk quasiparticles to become trapped.

The form we choose for the tunneling amplitude $\gamma(x)$ explicitly contains the
Aharonov-Bohm phase linear in $x$,
\begin{gather} 
\gamma(x)=\frac{\Gamma}{\sqrt{\pi}\ell_B}e^{-\frac{x^2}{L^2}}
e^{i\frac{x}{L}\frac{e^*}{e\:\:}N_\Phi},
\qquad N_\Phi=\frac{WL}{\ell_B^2}.
\label{eq:gammax}
\end{gather}
Here $N_\Phi$ is $2\pi$ times the number of flux quanta enclosed in
the area $WL$; $\Gamma/\ell_B$ is a measure of the tunneling amplitude
strength per unit length, which is assumed to be small enough to
warrant the weak-tunneling approximation of linear response.
We included a Gaussian
envelope to provide a smooth cut-off scale at length $L$; the Gaussian
form simplifies the integration over $x$ and $y$. The exact form of
the cut-off is not important when $L$ is large, and this is the regime
we are interested in, because temperature will introduce another,
smaller, cut-off length-scale. [For the case when $L$ is not so large
(i.e., $L/\ell_B\sim 1$), the approximation to $\gamma(x)$ in
Eq.~(\ref{eq:gammax}) is less accurate in that the Aharonov-Bohm phase
should not be simply linear, but should contain a quadratic piece to
account for the funneling in and out of the tunneling region.]

\begin{figure*}
\includegraphics[width=6.24375in]{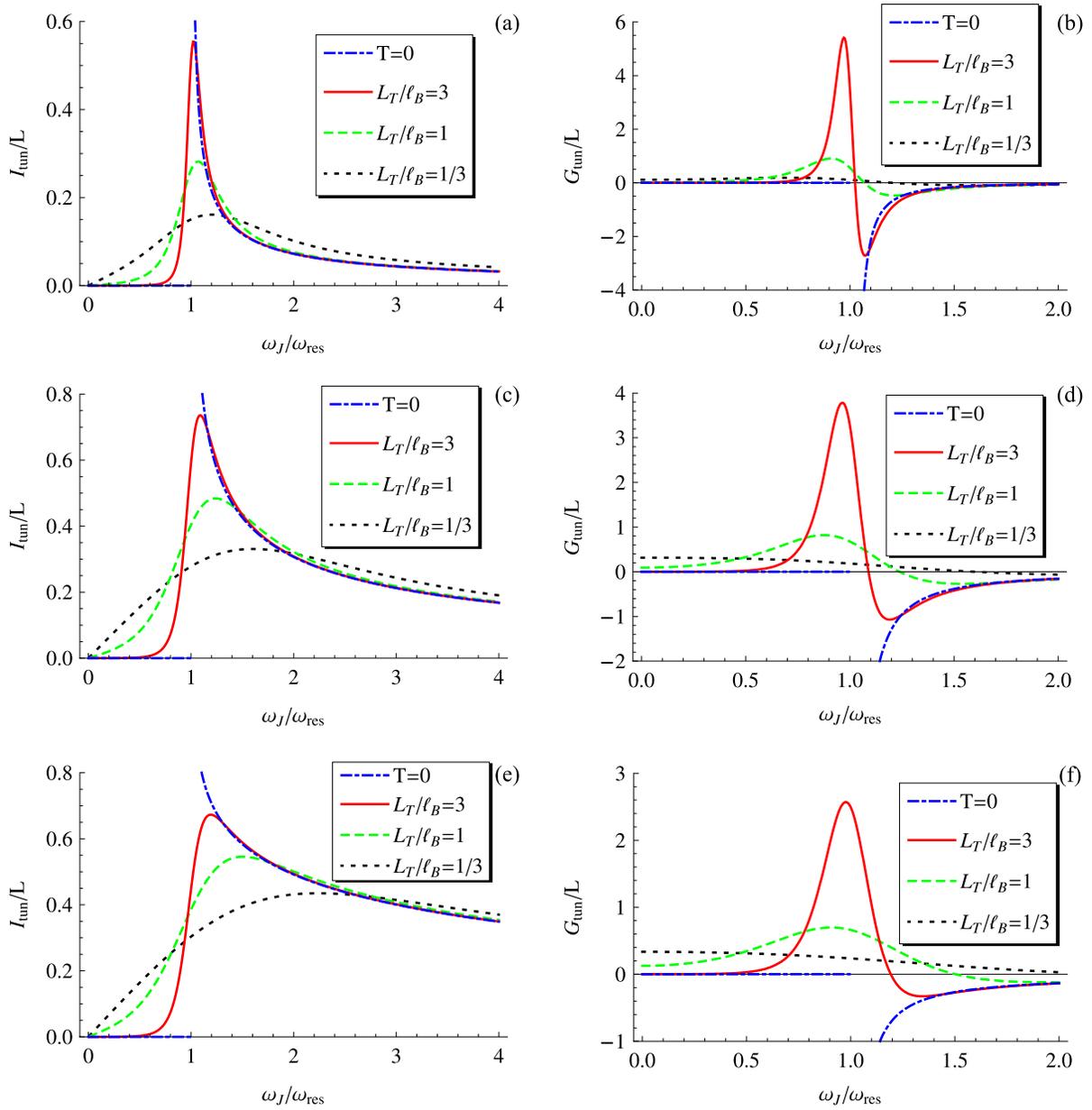}
\caption{
(Color online) Plots of the tunneling current per unit length (left column) and differential tunneling conductance per unit length (right column) for three states: the Pfaffian (a,b), the anti-Pfaffian (c,d) and the $\nu=\frac{2}{3}$ (e,f) state. The three states differ in their values for $e^*$, $g_c$ and $g_n$. Plotted are $I_\text{tun}/L$ and $G_\text{tun}/L$ as function of bias voltage at zero and finite temperatures.  At $T=0$ the current and conductance are zero for bias voltages below the threshold $\omega_J=\omega_\text{res}$ and diverge exactly at the resonance. At finite temperatures the divergence is reduced: the current reduces to a peak, the conductance reduces to a peak followed by a dip. When $L_T$, the length set by temperature, becomes smaller than $\ell_B$ the resonance becomes fully washed out and disappears. We set $|\Gamma|^2\omega_\text{res}^{2(g_c+g_n)-2}ee^*\equiv1$ and $W=10\ell_B$.
\label{fig:Gtunplots}
}
\end{figure*}

\begin{widetext}
One can carry out the integrals over $x$ and $y$ after recasting the
expression for the tunneling current Eq.~(\ref{eq:Itun-conntinuum}) in
terms of the (inverse) Fourier transforms of $P_g(t)$, 
\begin{gather}
\tilde P_g(\omega)\!=\!\begin{cases}
\displaystyle\theta(\omega)\;|\omega|^{2g-1}\frac{2\pi}{\Gamma(2g)}& 
\text{for $T=0$},\\
\displaystyle(2\pi T)^{2g-1}
\!B\left(g\!+\!i\frac{\omega}{2\pi T},g\!-\!i\frac{\omega}{2\pi T}\right)
e^{\frac{\omega}{2T}}& 
\text{for $T\ne0$},
\end{cases}
\end{gather}
where $\theta(s)$ is the Heaviside unit-step function and $B(a,b)$ is
the Euler beta function. 
In the limit $v_c\gg v_n$, i.e., neutral mode much slower than charged mode, the expression for the tunneling current becomes

\begin{equation}
\lim_{v_c\to\infty}I_\text{tun}=
e^*|\Gamma|^2\frac{L^2}{\ell_B^2}\int\frac{d\omega_1}{2\pi}\frac{d\omega_2}{2\pi}
\:\tilde P_{g_n\over 2}(\omega_1)\tilde P_{g_n\over 2}(\omega_2)
\tilde P_{g_c}(\omega_J-\omega_1-\omega_2)
\:e^{-\frac{1}{2}{N^*_\Phi}^2
\left[1-\frac{\omega_1-\omega_2}{\omega_\text{res}}\right]^2}
-(\omega_J\leftrightarrow -\omega_J),
\label{eq:Itunvcinf}
\end{equation}
where $N^*_\Phi\equiv (e^*/e) N_\Phi$ and $e^* V_j\equiv \omega_j$,
and $\omega_\text{res}\equiv e^*V_\text{res}$ is defined with respect to
the neutral velocity as in Eq.~(\ref{eq:res}).

Let us first consider Eq.~(\ref{eq:Itunvcinf}) in the limit of large
$L$, hence large $N^*_\Phi$, in which case the Gaussian in
Eq.~(\ref{eq:Itunvcinf}) reduces to a $\delta$-function that sets
$\omega_1-\omega_2=\omega_\text{res}$ (and a prefactor
$\sqrt{2\pi}\;\omega_{\text{res}}/N^*_\Phi$)%
;
in the limit of zero temperature one obtains
\begin{multline}
I_\text{tun}\to 
e|\Gamma|^2\frac{L}{W}
\;\text{sgn}(\omega_J)\:\:\theta(|\omega_J|-\omega_{\text{res}})
\:\:\omega_{\text{res}}^{2(g_c+g_n)-1}
\\
\!\!\!\!\!\!\!\!\!
\times
\frac{2^{-g_n}({2\pi})^{3/2}}{\Gamma(g_n)\Gamma(g_n+2g_c)}
\;
\left(\frac{|\omega_J|}{\omega_{\text{res}}}-1\right)^{2g_c+g_n-1}
F\left(1-g_n,g_n;2g_c+g_n;
\frac{1}{2}-\frac{1}{2}\:\frac{|\omega_J|}{\omega_{\text{res}}}\right)
\;,
\label{eq:ItunLinfTzero}
\end{multline}
\end{widetext}
where $F$ is the hypergeometric function. Notice the step function
$\theta(|\omega_J|-\omega_{\text{res}})$, so that, at $T=0$, the
current vanishes for biases below a threshold set by the
resonance. Near the resonance, the current scales as $I_\text{tun}\sim
\left(|\omega_J|/\omega_{\text{res}}-1\right)^{2g_c+g_n-1}$. At large
biases, far from the resonance, the current scales as
\begin{align}
I_\text{tun}\sim
\left(|\omega_J|/\omega_{\text{res}}\right)^{2g_c-1}
\begin{cases}
1&g_n<\frac{1}{2}\\
\ln |\omega_J|/\omega_{\text{res}}
& g_n=\frac{1}{2}\\
\left(|\omega_J|/\omega_{\text{res}}\right)^{2g_n-1}&g_n>\frac{1}{2}\\
\end{cases}.
\end{align}

Next, we consider Eq.~(\ref{eq:Itunvcinf}) for finite length $L$ and non-zero temperature $T$.
We find that either will smoothen the divergence at the resonance 
that exists for $T=0$ and $L\to\infty$.
Note that the ratio $I_\text{tun}/L$ is a useful quantity to compare different lengths $L$. 
 The
effect of finite temperature is remarkably similar to that of finite
length in the sense that we can define a length scale $L_T$ set by
temperature such that
\begin{gather}
\lim_{L\to\infty}\frac{1}{L}I_\text{tun}(L,T\ne 0)\simeq\frac{1}{L_T}I_\text{tun}(L_T,T=0)
\\
\frac{L_T}{\ell_B}\equiv\frac{e^* V_\text{res}}{2\pi T},\qquad
L_T=\frac{v_n}{2\pi T}\frac{e^*}{e}\frac{W}{\ell_B}.
\end{gather}
It was already emphasized by Bishara and Nayak~\cite{Bishara-Nayak}
for a two-point contact interferometer that $v_n/T$ sets a temperature
decoherence length scale; they define a temperature decoherence length
as $L_\phi=v_n/(2\pi T g_n)$ (for $v_c\to\infty$). Their definition
differs from ours by a factor of order one [since the two setups are different, exact comparison is not possible]. 


Plots of the tunneling current $I_\text{tun}$ and the differential tunneling conductance $G_\text{tun}=d I_\text{tun}/dV$ are shown in Fig.~\ref{fig:Gtunplots} for the following three quantum Hall states: the $\nu=\frac{5}{2}$ Pfaffian state ($e^*=e/4$, $g_c=1/8$, $g_n=1/8$), the $\nu=\frac{5}{2}$ anti-Pfaffian state ($e^*=e/4$, $g_c=1/8$, $g_n=3/8$), and the Abelian $\nu=\frac{2}{3}$ state ($e^*=e/3$, $g_c=1/6$, $g_n=1/2$). The current and conductance are plotted as a function of bias voltage and at different temperatures as indicated by $L_T$.  The tunneling current for a QLC is the main result of this paper, we plot the differential tunneling conductance as well because it is the conductance which is usually measured in experiment. 

Qualitatively the resonance at a QLC is independent of tunneling exponents $g_c$ and $g_n$, as the plots for the three different states in Fig.~\ref{fig:Gtunplots}  show more or less the same behavior:
at zero temperature the current and conductance are strictly zero below the resonance and diverge exactly at the resonance bias voltage of the QLC; at finite temperatures the resonance shows up as a strong peak in the current around the resonance bias voltage (strong peak followed by dip in the conductance) which becomes washed out if temperature becomes too high.
Note that  $I_\text{tun}(V)$ decays as power-law for $V\gg T$, so $G_\text{tun}$ will be negative here.  Qualitatively the resonance is a probe of a slow edge velocity. Quantitatively, the tunneling exponents $g_c$ and $g_n$ do affect the detailed shape of the resonance peak at finite temperature, and a precise observation of a resonance not only conveys information about the slow edge velocity but also about the tunneling exponents $g_c$ and $g_n$.%
\footnote{The quasiparticle \emph{charge} $e^*$ can be probed as well; however this requires a measurement of the tunneling current \emph{noise} in addition to a measurement of tunneling current (conductance).}

\section{accessible edge velocities\label{sec3}}

We would now like to address which range of slow edge
velocities can realistically be observed, and directly compare with the two-point contact interferometer setup  \cite{CFKSW9731, Bishara-Nayak}.
The lower bound is set by temperature (for both setups).  For the QLC, the 
 scale $L_T/\ell_B\approx 1$ is the cross-over region where
the resonance disappears. The lower bound $v_\text{min}^\text{QLC}$ on the slow edge velocity is then given by
\begin{equation}
v_\text{min}^\text{QLC}\simeq
\frac{2\pi}{(\frac{e^*}{e})(\frac{W}{\ell_B})}\frac{k_B T}{\hbar}\ell_B.
\end{equation}
For typical values, $T_\text{base}=10$mK, $\ell_B=10$nm, $W/\ell_B=10$, $e^*=e/3$, we find
$v_\text{min}^\text{QLC}\simeq 25$m/s. For the 2PC setup, the interference signal (which carries the edge velocity signature) is washed out when the spacing $x$ between the two contacts, i.e., the interferometer arm-length, is smaller than $L_\phi$. In current experiments, device fabrication limits $x\gtrsim 1\mu\text{m}$.  With $g_n=1/4$, this gives a lower bound of $v_\text{min}^\text{2PC}\simeq 2000 \text{m/s}$.  
Note that the QLC is sensitive to edge velocities up to two orders of magnitude slower compared to the 2PC setup. An intuitive explanation for this difference is to think of the QLC as an array of point contacts with a very small effective spacing $x$ which is much smaller than any spacing $x$ that can be fabricated for a 2PC setup.
 
For both the QLC and 2PC setups, the upper bound on the edge velocity that 
can be observed is given by the maximum voltage that can be applied to
the quantum Hall system without destroying it due to e.g. heating (a
current $I=V/R_H$ has to flow through the system). This maximum
voltage $V_\text{max}$ is not as clear-cut and may depend on sample,
specific experimental setup, and filling fraction. In terms of this
$V_\text{max}$ we have for the QLC setup
\begin{align}
v_\text{max}^\text{QLC}=\frac{1}{(\frac{W}{\ell_B})}\frac{e V_\text{max}}{\hbar}\ell_B.
\end{align}
To give a numerical estimate, for $eV_\text{max}=750 k_B
T_\text{base}$ one would find $v^\text{QLC}_\text{max}=1000$m/s.
The bulk excitation gap $T_\text{gap}$ likely sets the scale for $V_\text{max}$, but pre-factors are important [e.g. $eV_\text{max}\simeq T_\text{gap}$ and $e^* V_\text{max}\simeq2\pi T_\text{gap}$ differ by a factor 20]. For the 2PC setup our estimate gives $v_\text{max}^\text{2PC}\simeq  10^5\text{m/s}$ (for $x=1\mu\text{m}$).

\section{Conclusion\label{sec4}}

Given our estimates of the (non-overlapping) ranges of accessible edge velocities, we have to conclude that the QLC and 2PC setups complement each other quite well. A dedicated search for slow edge velocities should implement both setups in order to probe edge velocities from tens to ten-thousands of meters per second. Besides the different ranges of edge velocities, the main difference between the two setups is the signature of the edge velocity: for the QLC it is a \emph{resonance} in the tunneling conductance as function of bias, for the 2PC it is a \emph{modulation} of the interference signal within the tunneling conductance as function of bias;\cite{CFKSW9731, Bishara-Nayak} detecting a modulation in interference requires an extra experimental knob compared to detecting a resonance.

In this paper we assume the width $W$ of the QLC is constant, but disorder may lead to fluctuations of the width. As long as such fluctuations along the edge occur on scales larger than the magnetic lenght the resonance should survive, albeit with some broadening of the lineshape.
A feature at finite bias observed in device 2 of Ref.{~}{\onlinecite{Radu-etal}}, a channel-like geometry, can be due to a resonance, and leads us to expect that the proposed QLC setup is physically realizable.

In summary, we proposed and analyzed a device that can potentially
detect the presence 
of neutral edge modes at the edge of QH liquids,
by resolving velocities as small as tens of m/s. 
The ability to resolve these modes and measure their velocity of
propagation using a QLC  (possibly combined with a 2PC)  can provide a better quantitative
understanding of QH edge states, and can
help guide attempts to probe
quasiparticle statistics, both Abelian and non-Abelian. 

We thank M.~Kastner, J.~Miller, I.~Radu, and X.-G.~Wen, for
enlightening discussions. This work is supported in part by the DOE
Grant DE-FG02-06ER46316~(C.~C.)

\end{document}